\documentclass{amsart}
\usepackage{graphicx}

\numberwithin{equation}{section}

\newcommand{\vect}[1]{\boldsymbol{#1}}

\begin{document}

\title{Rotational elasticity and couplings to linear elasticity}

\author{C. G. B\"OHMER and N. TAMANINI}
\address{Department of Mathematics, University College London,\\
  Gower Street, London, WC1E 6BT, United Kingdom}
\email{c.boehmer@ucl.ac.uk}
\email{n.tamanini.11@ucl.ac.uk}

\subjclass[2000]{}
\date{\today}

\begin{abstract} 
It is the aim of the paper to present a new point of view on rotational elasticity in a nonlinear setting using orthogonal matrices. The proposed model, in the linear approximation, can be compared to the well known equilibrium equations of static linear elasticity. An appropriate kinetic energy is identified and we present a dynamical model of rotational elasticity. The propagation of elastic waves in such a medium is studied and we find two classes of waves, transversal rotational waves and longitudinal rotational waves, both of which are solutions of the nonlinear partial differential equations. For certain parameter choices, the transversal wave velocity can be greater than the longitudinal wave velocity. Moreover, parameter ranges are identified where the model describes an auxetic material. However, in all cases the potential energy functional is positive definite. Finally, we couple the rotational waves to linear elastic waves to study the behaviour of the coupled system. We find wave like solutions to the coupled equations and can visualise our results with the help of suitable figures.
\end{abstract}

\maketitle

\section{Introduction}

We are developing a theory of rotational elasticity based on using orthogonal matrices as the basic dynamical variables. Let us consider a three-dimensional elastic continuum, occupying an open connected set $\Omega \subset \mathbb{R}^3$ or the whole of $\mathbb{R}^3$, whose material points can only experience rotations and no displacements for now, they will be added later. Our Euclidean space is assumed to be equipped with Cartesian coordinates $\vect{x}=(x,y,z)$. To every point $\vect{x}\in\mathbb{R}^3$ we attach an orthonormal basis $\{\vect{e}_1,\vect{e}_2,\vect{e}_3\}$. In the initial state these basis vectors are assumed to be aligned with our Cartesian coordinate axes, i.e. $(\vect{e}_i)_{j}=\delta_{ij}$, but will become functions of $\vect{x}$ and time $t$ once the continuum is deformed. 

Models of this type have in fact a long tradition. One such model has first been introduced by MacCullagh in 1839, see~\cite{Whittaker_VolI}. He noted that it was not possible to describe optical phenomena by comparing the aether with an ordinary elastic solid. He thus introduced a new type of medium whose potential energy depended only on rotations. Similar models have been investigated by the Cosserat brothers which resulted in an extended framework of elasticity~\cite{Cosserat:1909}, often referred to as Cosserat elasticity. The main difference between Cosserat elasticity and classical elasticity is the assumed independence of displacements and rotations, often referred to as microrotations. When formulating an energy functional of classical elasticity, one assumes the integrand not to depend on the derivatives of rotation. It is well known that rotations about different axes do not commute, and thus we expect an inherently nonlinear theory, a similar investigation in a linearised setting was discussed in~\cite{na1,na2,na3}. The previous decade brought some revival of the study of Cosserat elasticity. The first existence result in the nonlinear setting~\cite{Neff:2006a} motivated various investigations which led to a variety of new results, see~\cite{Neff:2006b,Neff:2006c,Neff:2007a,Neff:2007b,Neff:2009,Neff:2009b,Neff:2010,Neff:2010a,Neff:2010b}. The existence result is based on a careful study of coercivity and an extended Korn inequality~\cite{Neff:2006a,Neff:2006c,Neff:2007a,Neff:2007b,Neff:2010}. 

Variants of the theory of Cosserat elasticity appear under various names in modern applied mathematics literature such as oriented medium, asymmetric elasticity, micropolar elasticity to mention a few popular ones, see e.g.~\cite{TT,gg1,gg2,gg3,Ericksen:1962a,Ericksen:1962b,Mindlin:1964,Toupin:1962,Toupin:1964,Green:1964,Eringen:1964a,Eringen:1964b,Schaefer:1967,sch2,sch3,Ericksen:1967,nowacki,capriz,Eringen99,dysz}. 

This field is also related to the theory of granular media, ferromagnetic materials, cracked media and liquid crystals to mention a few of them. There has also been an interest in the theory of Cosserat elasticity from a more theoretical physics point of view, see~\cite{Skyrme:1961,fhyo,Kat1,Kat2,Burnett:2008bx,Lazar:2009ga,Lazar:2010,Chervova:2010,Burnett:2012}. 

\section{Statement of the static problem}

In this Section and the next we are considering rotational deformations of the continuum and neglect displacements, these will be added in Section~\ref{sec4}. We assume that every material point can experience an independent rotation $\vect{e}^i \mapsto {\bf Q}(\vect{x})\vect{e}^i$, $i=1,2,3$, where ${\bf Q}(\vect{x})$ denotes the field of orthogonal matrices, ${\bf Q}^{\mathrm{T}}{\bf Q}={\bf Q}{\bf Q}^{\mathrm{T}}={\bf I}$ where ${\bf Q}^{\mathrm{T}}$ denotes the transpose and ${\bf I}$ the identity matrix. We consider orthogonal matrices with $\det {\bf Q} = +1$. The most general potential energy functional of such a static continuum is thus taken to be
\begin{align}
  V:=\int_{\Omega} f(\vect{x},{\bf Q},\partial{\bf Q},\partial^2{\bf Q},\ldots)\,dx\,dy\,dz
  \label{eqn_gen1}
\end{align}
where $\partial {\bf Q}$ ($\partial^2 {\bf Q}$, $\ldots$) denotes the first (second, $\ldots$) partial derivatives of ${\bf Q}$. The matrix-function ${\bf Q}(\vect{x})$ is the unknown quantity, our dynamical variable.

From a differential geometric point of view we are considering a Riemannian manifold whose metric is defined in the following way
\begin{align}
  g_{ij} = \delta_{mn} {\bf Q}_{mi} {\bf Q}_{nj}.
\end{align}
Here $i$ and $j$ are coordinate (an-holonomic) indices while the indices $m$ and $n$ denote the tangent-space (holonomic) or frame. In other words, we view the orthogonal matrix ${\bf Q}_{mi}$ as the tetrad of the manifold, see also~\cite{Boehmer:2012}. We made the choice in ${\bf Q}_{mi}$ that the first index $m$ denotes the frame and second index $i$ denotes the coordinate. Our manifold is considered to be globally flat and thus the manifold and its tangent space will both be isomorphic to $\mathbb{R}^3$. Thus, we do not have to distinguish the two types of indices explicitly in most of what follows, except when discussing left-invariance and right-invariance since both concepts operate in different spaces. 

\subsection{Basic equations of the static model}

The general framework (\ref{eqn_gen1}) of our rotational elasticity model is very wide. Hence, as a first step we identify a suitable subclass of variational functionals which is of mechanical and physical interest, and which is invariant under rigid rotations. This means the mapping ${\bf Q} \mapsto {\bf \bar{O}} {\bf Q}$, where ${\bf \bar{O}}$ is a constant orthogonal matrix, leaves the variational functional invariant. This means we use multiplication from the left to define invariance under rigid rotations. From the differential geometric point of view, we assume that rigid rotations of the frame (the first index of {\bf Q}, thus multiplication from the left) leave the variational functional invariant. This is referred to as frame-indifference.

Possible deformations of our elastic medium are characterised by the quantity 
\begin{align}
  {\bf K}_{ijk} := {\bf Q}_{mi} \partial_{j} {\bf Q}_{mk} 
  \label{defK}
\end{align}
where we sum over repeated indices. We could also write this as ${\bf K}_{j} := {\bf Q}^{\mathrm{T}} \partial_{j} {\bf Q}$, thereby suppressing the matrix indices, see also~\cite{Neff:2008n}. Had we defined our invariance under rigid rotations by multiplication from the right, we would have to consider the object $ {\bf Q} \partial_{j} {\bf Q}^{\mathrm{T}}$ instead. 

Let us next consider right-multiplications ${\bf Q} \mapsto {\bf Q} {\bf \bar{O}}$ or ${\bf Q}_{mi} \mapsto {\bf Q}_{mj} {\bf \bar{O}}_{ji}$. In this case the rigid rotation ${\bf \bar{O}}$ acts on the coordinate index (the second, thus multiplication from the right). If the coordinates are rotated, we must also take into account that $\vect{x} \mapsto {\bf \bar{O}} \vect{x}$ which gives $d\vect{x} \mapsto {\bf \bar{O}} d\vect{x}$ and $\nabla \mapsto {\bf \bar{O}}^{-1} \nabla = {\bf \bar{O}}^{\mathrm{T}} \nabla$.

Then the tensor ${\bf K}$ transforms as follows
\begin{align}
  {\bf K}_{ijk} &\mapsto ({\bf Q}_{mr}{\bf \bar{O}}_{ri}) {\bf \bar{O}}_{tj} \partial_{t} ({\bf Q}_{ms}{\bf \bar{O}}_{sk}) 
  \nonumber \\
  &= {\bf \bar{O}}_{ri} {\bf \bar{O}}_{tj} {\bf \bar{O}}_{sk} ({\bf Q}_{mr} \partial_{t} {\bf Q}_{ms})
  \nonumber \\
  &= {\bf \bar{O}}_{ri} {\bf \bar{O}}_{tj} {\bf \bar{O}}_{sk} {\bf K}_{rts}.
  \label{Ktrans}
\end{align}
Therefore, we see that the tensor ${\bf K}$ is frame-indifferent and transforms covariantly under rigid rotation from the right, see also~\cite{Neff:2008,Neff:2009ca}. 

In the context of Cosserat elasticity, this quantity ${\bf K}_{ijk}$ is sometimes referred to as contortion~\cite{Lazar:2009ga,Lazar:2010}, see also~\cite{Yavari:2012,Yavari:2012a}. This particular form of the contortion tensor gives rise to a globally flat or teleparallel space, see for instance~\cite{fhyo,Del:2013}. Note that ${\bf K}_{ijk}$ is, by definition, skew-symmetric in the first and third index, ${\bf K}_{ijk}=-{\bf K}_{kji}$. This follows from the fact that $\partial_{j} ({\bf Q}^{\mathrm{T}} {\bf Q}) = (\partial_{j} {\bf Q}^{\mathrm{T}}) {\bf Q} + {\bf Q}^{\mathrm{T}} \partial_{j} {\bf Q} = 0$. Therefore, ${\bf K}$ has 9 independent components.

The tensor ${\bf K}$ is related to the matrix ${\bf Q}^{\mathrm{T}} \mathrm{Curl} {\bf Q}$ which is of importance in the context of Cosserat elasticity, see for instance~\cite{Neff:2008n,Neff:2009}. Their relation is easy to derive, one finds
\begin{align}
  ({\bf Q}^{\mathrm{T}} \mathrm{Curl} {\bf Q})_{ij} = \varepsilon_{jrs} {\bf K}_{irs}.
\end{align}

In order to construct a suitable energy functional based on ${\bf K}$, it is useful to decompose it into its three irreducible components. There are only three irreducible pieces since ${\bf K}$ is skew-symmetric in one pair of indices. We can take the trace over two indices which gives us a vector, or we can contract ${\bf K}$ with the Levi-Civita symbol which gives a pseudo-scalar
\begin{align*}
  {\bf V}_{i} &:= {\bf K}_{jji} \\
  {\bf A} &:= \frac{1}{2} \varepsilon_{ijk} {\bf K}_{ijk}.
\end{align*}
One can verify that ${\bf V}$ and ${\bf A}$ have the correct transformation properties due to~(\ref{Ktrans}), this means ${\bf V}$ transforms like a vector under rigid rotation from the right and ${\bf A}$ like a pseudo-scalar.

Using the above, we arrive at
\begin{alignat*}{2}
  {\bf K}_{ijk}^{\!(1)} &:= \frac{1}{3} 
  \varepsilon_{ijk} {\bf A}
  &\qquad &\mbox{1 component}\\
  {\bf K}_{ijk}^{\!(2)} &:= \frac{1}{2}
  \left({\bf V}_{k}{\bf I}_{ij} - {\bf V}_{i}{\bf I}_{kj}\right) 
  &\qquad &\mbox{3 components}\\
  {\bf K}_{ijk}^{\!(3)} &:= {\bf K}_{ijk} - {\bf K}_{ijk}^{\!(1)} - {\bf K}_{ijk}^{\!(2)}
  &\qquad &\mbox{5 components}
\end{alignat*}
which is related the classical Cartan decomposition of $\mathrm{GL}(3)$. This follows from the observation that we can view $\varepsilon_{imn} {\bf K}_{mjn}$ as a $3 \times 3$ matrix.

The piece ${\bf K}_{ijk}^{\!(3)}$, which has 5 independent components, is closely related to the so-called $Q$-tensor of the Landau-de Gennes theory of uniaxial nematic liquid crystals, see~\cite{Ball:2008,Majumdar:2010}. The $Q$-tensor is symmetric and traceless, thereby having 5 independent components. However, in addition the $Q$-tensor has two equal eigenvalues. The elastic energy density in this theory is based on elastic invariants constructed from the irreducible components of the first partial derivatives of $Q$. In our model this corresponds to $\partial^2 {\bf Q}$, the second partial derivatives of ${\bf Q}$. Therefore, our general potential energy~(\ref{eqn_gen1}) is sufficiently wide to cover a large variety of materials.

A natural starting point would be to consider a functional of the type
\begin{align}
  V_1=\int_{\Omega} \left[c_1\|{\bf K}^{\!(1)}\|^2 + c_2\|{\bf K}^{\!(2)}\|^2 +
  c_3\|{\bf K}^{\!(3)}\|^2\right] \,dx\,dy\,dz
  \label{neweqn}
\end{align}
where $c_1$, $c_2$ and $c_3$ are positive constants and it would be tempting to regard those as elastic moduli. However, following the notation of~\cite{Neff:2010} we will refer to these as torsion moduli, opposed to curvature moduli to emphasise their geometric origin in our approach. By construction, the irreducible components of ${\bf K}$ are left-invariant and due to~(\ref{Ktrans}) transform covariantly. Therefore, the energy functional~(\ref{neweqn}) based on quadratic combinations of irreducible components is right-invariant.

A similar functional was considered in~\cite{Skyrme:1961} to formulate a unified field theory for mesons. Moreover, we denote the norm by $\|{\bf K}^{(a)}\|^2 = {\bf K}_{ijk}^{(a)} {\bf K}_{ijk}^{(a)}$ where $a=1,2,3$ and no summation over $a$. Note that a similar functional has been considered in the context of Cosserat elasticity~\cite{Neff:2010}. 

The term $\|{\bf K}^{\!(1)}\|^2$ deserves some special attention. This is precisely the object considered in~\cite{Chervova:2010,Burnett:2012} which eventually gives rise to the Dirac equation. The main motivation to consider this term is its conformal invariance, the Dirac equation is well-known to be invariant under conformal transformations. In this context a conformal transformation is a mapping $g_{ik} \mapsto \Omega g_{ij}$ where $g_{ij}$ is the metric of the (curved) Riemannian manifold and $\Omega$ is an arbitrary scalar function. One can argue that in the approach used in this paper, the irreducible pieces of the (Hodge dual) contortion tensor appear naturally. Therefore, if one aims to work with a conformally invariant object, $\|{\bf K}^{\!(1)}\|^2$ is the natural choice.

On the other hand, a conformally invariant curvature energy in the context of Cosserat elasticity is also of great interest and has been studied in~\cite{Neff:2009,Neff:2010,Neff:2010b}. In this case it turns out that the term $\|{\bf K}^{\!(3)}\|^2$ is invariant. Conformal invariance is less restrictive in this case as the entire formulation is based on three dimensional Euclidean space.

The three elastic moduli are not fully independent when one works in the whole of $\mathbb{R}^3$ because of the following identity. One can show that
\begin{align}
  -2 \|{\bf K}^{\!(1)}\|^2 - \|{\bf K}^{\!(2)}\|^2 + \|{\bf K}^{\!(3)}\|^2 = 
  4 \partial_i {\bf K}_{jji}^{\!(2)}.
  \label{eqn_div}
\end{align}
This identity has its roots in differential geometry, see Eq.~(5.9.18) in~\cite{mag} where it is expressed in terms of a rank 3 tensor. Of course, when $\Omega$ has a boundary, the above argument fails and the identity~(\ref{eqn_div}) leads to the appearance of boundary terms. One can regard ${\bf K}$ as an affine connection on a globally flat Riemannian manifold with curvature and torsion. Alternatively, one can work with the torsion tensor ${\bf T}$ directly~\cite{Boehmer:2010}, which is defined by $T_{jkl} := K_{jkl} - K_{jlk}$. The definition of the Riemann curvature tensor involves derivatives of the connection and contraction of the connection with itself. The vanishing of the Riemann curvature tensor is then equivalent to~(\ref{eqn_div}). Note that the vanishing of the Riemann curvature tensor is also often used in classical elasticity since one assumes that the distorted medium remains flat, thus implying certain compatibility conditions.

Next we can rewrite $\|{\bf K}^{\!(3)}\|^2$ in the energy functional $V_1$ in terms of the other two terms and a surface term which will not alter the resulting equations of motions. This results in the functional
\begin{align}
  V_2=\int_{\Omega} \left[\hat{c}_1\|{\bf K}^{\!(1)}\|^2 + \hat{c}_2\|{\bf K}^{\!(2)}\|^2 \right] dx\,dy\,dz
  \label{eqn_gen2}
\end{align}
where $\hat{c}_1 = c_1+2c_3$ and $\hat{c}_2 = c_2+c_3$.

\subsection{Basic equations of the linearised model}

It is known that every orthogonal matrix ${\bf Q}$ can be written as the exponential of a skew-symmetric matrix. It turns out to be convenient to analyse the nonlinear equations and their relation with linear elasticity by writing ${\bf Q} = \exp(\vect{\omega}^{\star})$. We denote by $\vect{\omega}^{\star}$ the skew-symmetric matrix dual to the vector $\vect{\omega}$. This means $(\vect{\omega}^{\star})_{ik} = \varepsilon_{ijk}\vect{\omega}_{j}$ or, written out explicitly,
\begin{align*}
  \vect{\omega} = \begin{pmatrix} \omega_x \\ \omega_y \\ \omega_z \end{pmatrix},\qquad
  \vect{\omega}^{\star} = \begin{pmatrix}
    0 & -\omega_z & \omega_y \\ \omega_z & 0 & -\omega_x \\ -\omega_y & \omega_x & 0
  \end{pmatrix}.
\end{align*}
In other words, rather than using elements of the group ${\bf Q} \in \mathrm{SO}(3)$ as the dynamical or state variables, we are using elements of its Lie algebra $\vect{\omega}^{\star} \in \mathfrak{so}(3)$, see~\cite{Neff:2010} where this approach is used in linear Cosserat elasticity. It should be noted that there exists a simple expression for the matrix ${\bf Q}$ which is sometimes referred to as Rodrigues' formula
\begin{align*}
  \exp(\vect{\omega}^{\star}) = {\bf I} + \sin(\|\vect{\omega}\|) \frac{\vect{\omega}^{\star}}{\|\vect{\omega}\|} + (1 - \cos(\|\vect{\omega}\|)) \frac{(\vect{\omega}^{\star})^2}{\|\vect{\omega}\|^2}.
\end{align*}

We can expand the orthogonal matrix in $\vect{\omega}$ and obtain
\begin{align}
  {\bf Q} = \exp(\vect{\omega}^{\star}) = {\bf I} + \vect{\omega}^{\star} + O(\|\vect{\omega}\|^2),
  \label{eqn_f}
\end{align}
where $\|\vect{\omega}\|^2$ denotes the standard vector norm $\|\vect{\omega}\|^2 = \omega_x^2 + \omega_y^2 + \omega_z^2$. This results in 
\begin{align*}
  {\bf K}_{ijk} &:= {\bf I}_{im} \partial_{j} {(\vect{\omega}^{\star})}_{km} + O(\|\vect{\omega}\|^2) \\
  &= \partial_{j} {(\vect{\omega}^{\star})}_{ki} + O(\|\vect{\omega}\|^2) \\
  &= \partial_{j} \varepsilon_{kmi} \vect{\omega}_m + O(\|\vect{\omega}\|^2)
\end{align*}
and therefore in the linearised setting we have
\begin{align*}
  {\bf K}_{ijk} = \varepsilon_{kmi} \partial_{j} \vect{\omega}_m 
  + O(\|\vect{\omega}\|^2).
\end{align*}
We should note that $\partial_{j} \vect{\omega}_m$ is nothing but the matrix $\nabla \vect{\omega}$.

In~\cite{Neff:2009,Neff:2010,Neff:2010b}, for instance, the rotational part of the Cosserat medium is modelled using the skew-symmetric infinitesimal microrotation $\bar{A} \in \mathfrak{so}(3)$ ($\vect{\omega}^{\star}$ in our notation) which is used to define the micropolar curvature tensor or curvature-twist tensor $\nabla \mathrm{axl} \bar{A}$ ($\varepsilon_{mjk}{\bf K}_{jnk}/2$ or $\nabla \vect{\omega}$ in our notation). This allows us to identify our coupling constants $c_i$ with the micropolar moduli of~\cite{Neff:2009,Neff:2010,Neff:2010b} as follows
\begin{align*}
  c_1 = \frac{3\alpha+\beta+\gamma}{4},\qquad
  c_2 = \frac{\gamma-\beta}{4},\qquad
  c_3 = \frac{\gamma+\beta}{4},
\end{align*}
provided we consider the linearised equations of our model.

Let us now assume $\vect{\omega}$ to be small and let us expand the terms in the functional~(\ref{neweqn}) up to terms $O(\|\vect{\omega}\|^3)$. This gives
\begin{align*}
  \|{\bf K}^{\!(1)}\|^2 &= \frac{2}{3} (\mathrm{div}\, \vect{\omega})^2 + 
  O(\|\vect{\omega}\|^3)\\
  \|{\bf K}^{\!(2)}\|^2 &= \|\mathrm{curl}\, \vect{\omega}\|^2 + O(\|\vect{\omega}\|^3)\\
  \|{\bf K}^{\!(3)}\|^2 &= -\|\mathrm{curl}\, \vect{\omega}\|^2 - \frac{2}{3} (\mathrm{div}\, \vect{\omega})^2 + 2 \partial_m \omega_n \partial_m \omega_n + O(\|\vect{\omega}\|^3).
\end{align*}
It is also useful to consider the right-hand side of~(\ref{eqn_div}) which becomes
\begin{align}
   2 \partial_i {\bf K}_{jji}^{\!(2)} = 
   \partial_m \omega_n \partial_m \omega_n - 
   (\mathrm{div}\, \vect{\omega})^2 -
   \|\mathrm{curl}\, \vect{\omega}\|^2 + 
   O(\|\vect{\omega}\|^3), 
   \label{eqn_div2}
\end{align}
which is consistent with the identity~(\ref{eqn_div}).

Therefore, in the linear approximation the energy functional~(\ref{eqn_gen2}) becomes
\begin{align}
  V_3 = \int_{\Omega} \alpha (\mathrm{div}\, \vect{\omega})^2 + 
  \beta \|\mathrm{curl}\, \vect{\omega}\|^2 \,dx\,dy\,dz
  \label{eqn_lin}
\end{align}
where we denoted $\alpha = 2 \hat{c}_1/3$ and $\beta = \hat{c}_2$. Comparison with Eq.~(3.9) of~\cite{Neff:2010} shows that necessary and sufficient conditions for strict Legendre-Hadamard ellipticity of~(\ref{eqn_gen2}) are simply $\alpha > 0$ and $\beta > 0$.

Variation of the quadratic functional~(\ref{eqn_lin}) with respect to our independent (dynamical) variable, the components of $\vect{\omega}^{\star}$, yields
\begin{align*}
  \delta V_3 = \int_\Omega \left(
  \frac{\partial V_3}{\partial \vect{\omega}^{\star}_{ab}}\delta \vect{\omega}^{\star}_{ab} + 
  \frac{\partial V_3}{\partial (\partial_c \vect{\omega}^{\star}_{ab})}\delta 
  (\partial_c \vect{\omega}^{\star}_{ab})\right) dx\,dy\,dz\,,
\end{align*}
which after integration by parts and using Gauss' theorem gives
\begin{align*}
  \delta V_3 = \int_\Omega \left(
  \frac{\partial V_3}{\partial \vect{\omega}^{\star}_{ab}} -
  \partial_c \frac{\partial V_3}{\partial (\partial_c \vect{\omega}^{\star}_{ab})} \right)
  \delta \vect{\omega}^{\star}_{ab}\,dx\,dy\,dz + \int_{\partial \Omega} 
  n_c \frac{\partial V_3}{\partial (\partial_c \vect{\omega}^{\star}_{ab})}\, 
  \delta (\partial_c \vect{\omega}^{\star}_{ab})\, dS.
\end{align*}
To simplify matters, we assume Dirichlet boundary conditions and henceforth neglect the boundary term. This gives us the Euler-Lagrange equations
\begin{align*}
  \alpha\, \mathrm{grad}(\mathrm{div}\, \vect{\omega}) - 
  \beta\, \mathrm{curl}\, \mathrm{curl}\, \vect{\omega} = 0.
\end{align*}
This system of equations has already been known by Cauchy and led to modifications due to MacCullagh and also Neumann, we refer the reader to~\cite{Whittaker_VolI}.

The vector identity $\mathrm{curl}\, \mathrm{curl}\, \vect{\omega} = \mathrm{grad}(\mathrm{div}\, \vect{\omega}) - \Delta \vect{\omega}$ enables us to rewrite the Euler-Lagrange equations in the form
\begin{align}
  \beta\, \Delta \vect{\omega} + (\alpha - \beta)\, 
  \mathrm{grad}(\mathrm{div}\, \vect{\omega}) = 0.
  \label{eqn_leom}
\end{align}
Comparing with the well-known equilibrium equations of linear elasticity $\mu\, \Delta \vect{\omega} + (\mu+\lambda)\, \mathrm{grad}(\mathrm{div}\, \vect{\omega}) = 0$, we would relate $\beta = \mu$ and $\alpha = \lambda + 2\mu$ to get a corresponding theory. This analogy is useful when trying to interpret the fully non-linear theory. Note that the equilibrium equations of linear elasticity are derived from the functional
\begin{align}
  \int_\Omega \left[ \frac{\lambda}{2} \|e_{ii}\|^2 + 
  \mu \|e_{ij}\|^2 \right] dx\,dy\,dz.
  \label{eqn_elas}
\end{align}
where we denote $e_{ij} = (\partial_i u_j + \partial_j u_i)/2$, see~\cite{Love:1944,Muskhelishvili:1953,Sokolnikoff:1956}.
It should be noted here that parts of the second term of the functional can be found in the right-hand side of the derivative term~(\ref{eqn_div2}). Another interesting fact of linear elasticity is that not all the irreducible pieces of the matrix $\partial_i u_j$ are regarded as the fundamental building blocks of the potential energy functional. The skew-symmetric part, namely the (macro) rotations, are in general not part of the potential energy. It is the aforementioned Cosserat theory of elasticity which incorporates local rotations of material points.

The functional~(\ref{eqn_elas}) is obviously positive definite if the elastic moduli $\lambda$ and $\mu$ are assumed to be positive. Similarly, the functional~(\ref{eqn_lin}) is also positive definite provided that $\alpha$ and $\beta$ are positive. Therefore, when comparing the nonlinear functional~(\ref{eqn_gen2}), which is positive definite provided $c_i > 0$ for $i=1,2,3$, with the linear elasticity functional~(\ref{eqn_elas}), we are led to identify
\begin{align*}
  \lambda = 2\,\left(\frac{1}{3}c_1 - c_2 - \frac{1}{3}c_3\right),\qquad
  \mu = 2\,(c_2 + c_3).
\end{align*}

This immediately suggests that Poisson's ratio for our rotational medium may not be between $-1$ and $1/2$ and, moreover, that the transverse wave velocity may be greater than the longitudinal wave velocity. By straightforward comparison we find the rotational analogues for Poisson's ratio $\nu$ and Young's modulus $E$, respectively
\begin{align*}
  &\nu = \frac{\lambda}{2(\lambda+\mu)} =
  \frac{c_1-3c_2-c_3}{2c_1-3c_2+c_3},\\
  &E = \frac{\mu(3\lambda+2\mu)}{\lambda+\mu} = 
  \frac{6(c_1-2c_2)(c_2+c_3)}{2c_1-3c_2+c_3},
\end{align*}
see also~\cite{Neff:2010}. Thus, by choosing the constants $c_i>0$ appropriately, it is possible to define materials with negative Poisson's ratio, also known as auxetic materials, see~\cite{Lakes:1987}. For ordinary materials with positive Young's modulus $E>0$ and positive Poisson's ratio in the range $\nu \in (0,1/2)$, our positive constants have to satisfy $c_1 > 3c_2 + c_3$. An auxetic material with $\nu \in (-1,0)$ is realised when $2c_2 < c_1 < 3c_2 + c_3$.

Moreover, the constants $c_i>0$ can also be chosen so that the analogous linearised functional~(\ref{eqn_elas}) is not positive definite while the nonlinear theory in fact is. This fact is important when stability is investigated.

The equilibrium equations of our rotational elasticity theory are derived from an energy functional. Therefore, the nonlinear model is automatically a quasilinear, second order PDE system in divergence form~\cite{Evans:1998}. A natural topic to address is the convexity of the potential energy. For the linearised model~(\ref{eqn_lin}) one can immediately verify that the functional is coercive, independent of $\vect{\omega}$ and convex in $\partial\vect{\omega}$. Thus, existence and uniqueness of the linearised Dirichlet boundary value problem are established in appropriate function spaces. In the nonlinear setting, the issue of existence and uniqueness is quite subtle and requires a careful investigation. Existence results in nonlinear Cosserat elasticity can be found in~\cite{Neff:2006a,Neff:2006c,Neff:2007a,Neff:2007b,Neff:2010}.

\subsection{Identifying kinetic energy}

So far our model of rotational elasticity has been static. The next logical step is to introduce kinetic energy into the energy functional~(\ref{eqn_gen2}). To do this, we add a term of the form $\|\dot{{\bf Q}}\|^2$ to the energy functional. To be consistent with our previous notation, we have
\begin{align}
  \|\dot{{\bf Q}}\|^2 = {\rm trace} (\dot{\bf Q} \dot{{\bf Q}}^{\mathrm{T}})
\end{align}
which is invariant under rigid rotations. If we now linearise this form of kinetic energy according to Eq.~(\ref{eqn_f}), we find
\begin{align}
  \|\dot{{\bf Q}}\|^2 = 2 \|\dot{\vect{\omega}} \|^2 + O(\|\vect{\omega}\|^3).
\end{align}
Note that one regards $\|\dot{{\bf Q}}\|$ as angular speed (modulus of the angular velocity vector) and $\rho \|\dot{{\bf Q}}\|^2$ as rotational energy or angular kinetic energy, where $\rho$ is the energy density of the medium.

\subsection{Statement of the full problem}
\label{subsec:full}

After identifying the kinetic energy, we can now formulate the complete variational functional of rotational elasticity in our setting. 

Let us find a skew-symmetric matrix $\vect{\omega}^{\star}$ such that ${\bf Q}=\exp(\vect{\omega}^{\star})$ the variational functional or action is stationary 
\begin{align}
  V = \int_{\Omega_T} \left[c_1\|{\bf K}^{\!(1)}\|^2 + c_2\|{\bf K}^{\!(2)}\|^2 + c_3\|{\bf K}^{\!(3)}\|^2 - \rho \|\dot{{\bf Q}}\|^2\right] dx\,dy\,dz\,dt
  \label{eqn_full}
\end{align}
where $\Omega_T := \Omega \times (0,T]$ with $T>0$. We assume Dirichlet boundary conditions, namely ${\bf Q}={\bf I}$ on $\partial\Omega$. When $\Omega$ is the whole of $\mathbb{R}^3$ we assume ${\bf Q} \rightarrow {\bf I}$ as $|\vect{x}| \rightarrow \infty$, which is equivalent to assuming that the skew-symmetric matrix $\vect{\omega}^{\star} \rightarrow {\bf 0}$ at the boundary. In certain situations appropriate radiation conditions have to be taken into account. 

At this point it worth emphasising some important differences between the current work and~\cite{Boehmer:2010}. In the latter, the entire Lagrangian is multiplied by a density, thereby working with a fluid rather than an elastic medium, see (4.18) in~\cite{Boehmer:2010}. This density is also regarded as a dynamical degree of freedom. Thus, variations with respect to this density will force the Lagrangian to be zero as it has the same effect as a Lagrange multiplier. Therefore, the full problem~(\ref{eqn_full}) is a dynamically distinct model, which has also been studied in~\cite{Boehmer:2012} where it was shown that soliton-type solutions exist and that these have a topological origin. 

\section{Propagation of rotational waves}

\subsection{Assumptions of the model}

We are now discussing solutions to the fully nonlinear problem. The explicit equations and very complicated expressions. However, it is possible to state those equations symbolically, see Section 2~(d) of~\cite{Boehmer:2012}. It turns out that these equations can be simplified considerably by a variety of assumptions which we will explain in the following.

As a first simplifying assumption, we consider a medium which can only experience rotation about one axis, the $z$-axis say. This means we can assume
\begin{align}
  \vect{\omega}^{\star} &= \varphi \begin{pmatrix}
    0 & 1 & 0 \\ -1& 0 & 0 \\ 0 & 0 & 0 \end{pmatrix},
  \nonumber \\
  {\bf Q} &= \exp(\vect{\omega}^{\star}) = \begin{pmatrix}
    \cos\varphi & -\sin\varphi & 0 \\ 
    \sin\varphi & \cos\varphi & 0 \\ 
    0 & 0 & 1 \end{pmatrix},
\end{align}
where $\varphi=\varphi(x,y,z,t)$ is an arbitrary function of the three spatial variables and time. This choice for $\vect{\omega}^{\star}$ means that $\omega_x \equiv 0$ and $\omega_y \equiv 0$ in Eq.~(\ref{eqn_f}) and thus $\mathrm{div} \vect{\omega} = \partial_{x} \omega_{z}$. To further simply the problem, we follow the well-known approach of classical elasticity~\cite{Love:1944}. Note the interesting twist waves in liquid crystals analysed in~\cite{Ericksen:1968}.

Transversal rotational waves: We assume that the medium is homogeneous along the single axis of rotation, this means the $z$-axis. Thus, we restrict ourselves to $\varphi=\phi_{\rm t}(x,y,t)$, which corresponds to the choice $\mathrm{div}\, \vect{\omega} \equiv 0$. 

Longitudinal rotational waves: By choosing $\varphi$ to depend only on the $z$-coordinate and time, $\varphi=\phi_{\rm l}(z,t)$, we effectively set $\mathrm{curl}\, \vect{u} \equiv 0$.

We will use this approach to find rotational waves propagating through the elastic medium.

\subsection{The Helmholtz equation}

In general, since the matrix $\vect{\omega}^{\star}$ has three independent components, the variation with respect to its components yields three coupled second order PDEs. Using the above ansatz for the transversal and longitudinal rotational waves, two of the three equations are identically satisfied, respectively, and the remaining equations become
\begin{align*}
  \frac{1}{2}(c_2+c_3)(\partial_{xx} + \partial_{zz})\phi_{\rm t} - 
  \rho \partial_{tt} \phi_{\rm t} &= 0 \qquad \mathrm{transversal}\\
  \frac{1}{3}(c_1+2c_3) \partial_{zz} \phi_{\rm l} - 
  \rho \partial_{tt} \phi_{\rm l} &= 0 \qquad \mathrm{longitudinal}
\end{align*}
respectively. It is remarkable that these equations are linear, however, we did not linearise the full equations at any point. It is our ansatz that makes the nonlinear terms disappear.

Therefore, in our elastic medium we find two types of rotational waves. Due to the nonlinear nature of the equations $\phi = \phi_{\rm t} + \phi_{\rm l}$ is not a solution of the Euler-Lagrange equations. As expected, the superposition principle does not hold.

The transversal rotational waves travel in the $x-y$ planes and are homogeneous along the $z$-axis while the longitudinal rotational waves travel along the $z$-axis only. The corresponding wave velocities are given by
\begin{align}
  v_{\rm t} = \sqrt{\frac{c_2+c_3}{2\rho}},\qquad
  v_{\rm l} = \sqrt{\frac{c_1+2c_3}{3\rho}}
\end{align}
with their ratio given by
\begin{align}
  \nu = \frac{v_{\rm t}}{v_{\rm l}} = \sqrt{\frac{3}{2}\frac{c_2+c_3}{c_1+2c_3}}.
  \label{eqn_nu}
\end{align}
The requirement of real wave speeds implies the conditions $c_2+c_3 > 0$ and $c_1+2c_3 > 0$, see~\cite{Neff:2010} and the remark after Eq.~(\ref{eqn_gen2}).

It is a well known fact in classical elasticity that the wave velocity of the longitudinal waves is always greater than the transversal wave velocity. This does not hold in our rotational elasticity model. For an ordinary material we found $c_1 > 3c_2 + c_3$. Inserting this into~(\ref{eqn_nu}) the ratio of the velocities becomes
\begin{align}
  \nu = \sqrt{\frac{3}{2}\frac{c_2+c_3}{c_1+2c_3}} < \sqrt{\frac{1}{2}}
\end{align}
which is in agreement with the corresponding equations of classical elasticity. On the other hand, for an auxetic material $2c_2 < c_1 < 3c_2 + c_3$, and thus
\begin{align}
  \sqrt{\frac{1}{2}} < \nu < \sqrt{\frac{3}{4}}.
\end{align}
These considerations show that our rotational elasticity model shares many features with well know classical elasticity, however, it also contains many new interesting features.

\subsection{Visualising transversal rotational waves assuming planar symmetry}

If we are now separating variables and assume
\begin{align}
  \phi_t(x,y,t) = \cos(\omega t) v(x,y)
\end{align}
we obtain
\begin{align}
  (\Delta + k^2) v = 0,\qquad
  k = \omega \sqrt{\frac{2\rho}{c_2+c_3}}
\end{align}
the Helmholtz equation and the dispersion relation for the medium. The Helmholtz equation is a well understood equation within the theory of PDEs. However, we will further simplify our model.

Note that, as expected, the function $\phi \equiv \mathrm{const.}$ solves these equations and hence it is natural to regard this solution as the ground state where all material points of the medium are aligned.

In addition to assuming rotations about the $z$-axis only and homogeneity along the axis of rotation, we will now also specify a point in the plane orthogonal to the axis of rotation, the $x-y$ plane. This means, we are assuming rotational symmetry in the $x-y$ plane, we write $\phi(x,y,z) = e^{i \omega t}v(r)$ with $r^2=x^2+y^2$. The nonlinear problem now reduces to a single ordinary differential equation which is given by
\begin{align}
  v''(r) + \frac{1}{r}v'(r) + \frac{2 \rho\, \omega^2}{c_2+c_3}v(r) = 0,
\end{align}
and is solved by a linear combination of Bessel functions of first and second kind. We require the solution to be regular at $r=0$, therefore choosing one constant of integration to eliminate the Bessel function of second kind. We are left with the solution
\begin{align}
  v(r) = v_0\, \mathcal{J}_0 \left(2\omega\sqrt{\frac{\rho}{c_2+c_3}}\,r\right)
\end{align}
where $v_0 = v(0)$ and $\mathcal{J}_0$ denotes the Bessel function of first kind.

\begin{figure}[!ht]
\begin{center}
\includegraphics[width=.48\textwidth]{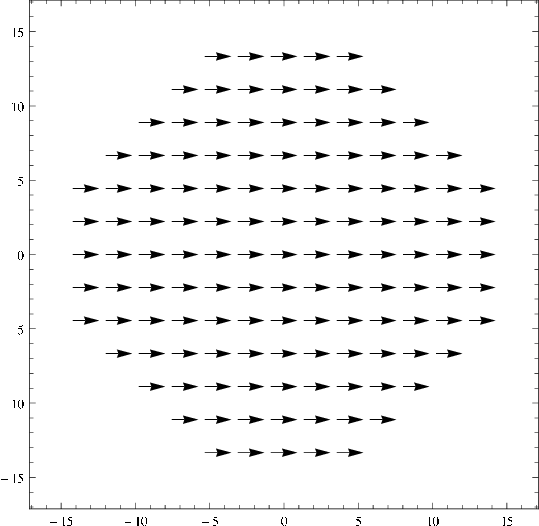}
\hfill
\includegraphics[width=.48\textwidth]{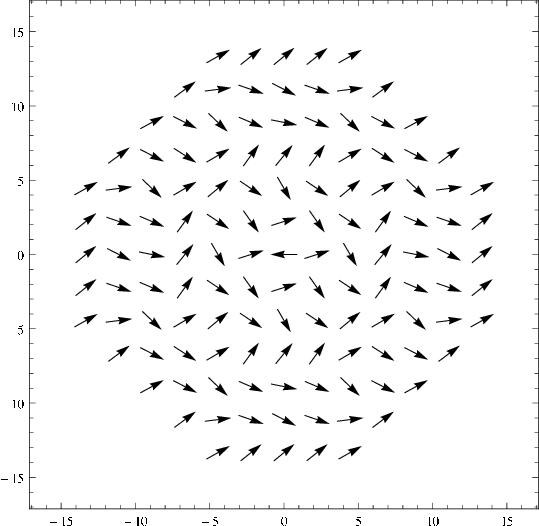}
\caption{Rotational elasticity visualised. The left panel shows the undistorted medium where all material points are aligned parallelly to the $x$-axis. The right panel depicts the solution of the transversal rotational wave equation when the central material point is rotated by a half turn.}
\label{fig1}
\end{center}
\end{figure}

This solution of rotational elasticity can now be visualised in the following manner. Let us attach to every material point in the medium an arrow to indicate its orientation. This is the practical realisation of attaching an orthonormal basis to every point in space. The choice $\phi \equiv 0$ corresponds to the situation where all arrows are parallel to the $x$-axis, which means that the medium under consideration is assumed to be aligned. We choose $\omega\sqrt{2\rho/(c_2+c_3)}=1$ and $\omega_0=\pi$, where the latter means that we rotate the central material point by a half turn. Thus we investigate the effect of the resulting rotational wave on the rest of the medium.  The medium in its ground state $\phi \equiv 0$ in shown in the left panel of Fig.~\ref{fig1} while the distorted medium is shown in the right panel.

\section{Interaction of elastic and rotational waves}
\label{sec4}

The propagation of rotational waves shows some similarities with the well-know propagation of elastic waves~\cite{Love:1944}. In the following we will consider the coupling of both waves types. Recall that the energy functional for linear elasticity is given by 
\begin{align}
  V_{\rm elastic} = \int_{\Omega_T} \left[ \frac{\lambda}{2} (e_{ii})^2 + 
  \mu \|e_{ij}\|^2 - \frac{\rho}{2} \|\dot{\vect{u}}\|^2 \right] dx\,dy\,dz\,dt
  \label{eqn_elas_full}
\end{align}
where $e_{ij} = (\partial_i u_j + \partial_j u_i)/2$ and $\vect{u}$ is the usual displacement vector. For concreteness we repeat~(\ref{eqn_full})
\begin{align}
  V_{\rm rotational} = \int_{\Omega_T} \left[c_1\|{\bf K}^{\!(1)}\|^2 + c_2\|{\bf K}^{\!(2)}\|^2 + c_3\|{\bf K}^{\!(3)}\|^2 - \rho_{\rm rot} \|\dot{{\bf Q}}\|^2\right] dx\,dy\,dz\,dt.
  \label{eqn_full_rot}
\end{align}
Wave propagation in unbounded continua with microstructure has also been studied recently in~\cite{Madeo:2013a,Madeo:2013b}

\subsection{Interaction terms and the full coupled problem}

In order to propose meaningful interaction terms, it makes sense to only combine terms of the same irreducible spaces. The energy functional for linear elasticity~(\ref{eqn_elas_full}) contains as irreducible pieces, a trace $e_{ii}$ and a trace-free symmetric rank 2 tensor $e_{ij} - 1/3 e_{kk} \delta_{ij}$. Therefore, these terms can be coupled to the terms ${\bf K}^{\!(1)}$ and ${\bf K}^{\!(3)}$ of the rotational energy functional~(\ref{eqn_elas_full}), respectively. The energy functional for the interaction is thus given by
\begin{multline} 
  V_{\rm interaction} = \int_{\Omega_T} \Bigl[\chi_1 \varepsilon_{ijk} {\bf K}^{\!(1)}_{ijk} e_{mm} \\+ \chi_3 \varepsilon_{ikn} {\bf K}^{\!(3)}_{kjn} (e_{ij} - 1/3 e_{kk} \delta_{ij}) \Bigr] dx\,dy\,dz\,dt
  \label{eqn_int}
\end{multline}
where we neglected all possible interaction terms of higher than second order. 

In analogy to Section \ref{subsec:full} we can now formulate the complete variational problem in our setting. Let us find a vector $\vect{u}$ and a skew-symmetric matrix $\vect{\omega}^{\star}$ such that $\vect{u}$ and ${\bf Q}=\exp(\vect{\omega}^{\star})$ the variational functional or action is stationary
\begin{align}
  V = V_{\rm elastic} + V_{\rm rotational} + V_{\rm interaction}
  \label{eqn_full_with_int}
\end{align}
with appropriate boundary conditions, and possibly taken into account radiation conditions when needed. Since the functional~(\ref{eqn_full_with_int}) contains six dynamical degrees of freedom, there are in total six coupled nonlinear Euler-Lagrange equations.

We explicitly constructed some plane wave solutions of the rotational waves and we know that similar solutions exist for elastic waves. As the full set of equations is very complicated, we will restrict our investigation to situations where both types of waves are either transversal or longitudinal. This allows us to construct explicit solutions which contain some very interesting features.

Let us briefly comment on the possible coupling term
\begin{align*}
  \frac{\mu_c}{2} \|\mathrm{curl}\, \vect{u} - 2\vect{\omega}\|^2
\end{align*}
where $\mu_c$ is the so-called Cosserat couple modulus, see~\cite{Neff:2006c,Neff:2007a,Neff:2013a}. In our approach to Cosserat elasticity, this term is rather unnatural. It would involve the quantity ${\bf Q}$ ($\vect{\omega}$ when linearised) directly, instead of its derivatives which we use as the building blocks of the model. Moreover, our basic quantity ${\bf A}$ was chosen to be invariant under rigid rotations. Therefore any coupling term involving ${\bf Q}$ has to be constructed very carefully in order to ensure its invariance.

\subsection{Transversal--Transversal coupling}

Let us begin with assuming that the elastic medium can only experience rotation about the $z$-axis. Moreover, we assume that the medium is homogeneous along this axis of rotation, this means we can write $\omega_x \equiv 0$, $\omega_y \equiv 0$, and $\omega_z = \phi(x,y,t)$. Analogously, we assume that our medium can only experience displacements along the axis of rotation. Homogeneity of those displacements implies that $u_x \equiv 0$, $u_y \equiv 0$, and $u_z = \psi(x,y,t)$. One easily verifies that indeed $\mathrm{div} \vect{\omega} = 0$ and $\mathrm{div} \vect{u} = 0$. These assumptions simplify the Euler-Lagrange equations considerably. Four of the six equations are identically satisfied and one is left with two equations
\begin{align}
  \frac{1}{2}(c_2+c_3) (\partial_{xx} + \partial_{yy}) \phi - \frac{1}{4} \chi_3 (\partial_{xx} + \partial_{yy}) \psi - \rho_{\rm rot} \partial_{tt} \phi &= 0 \\
  \mu (\partial_{xx} + \partial_{yy}) \psi - \chi_3 (\partial_{xx} + \partial_{yy}) \phi - \rho \partial_{tt} \psi &= 0.
  \label{eqn_tt1}
\end{align}
The absence of the coupling constant $\chi_1$ is expected as the corresponding term in~(\ref{eqn_int}) identically vanishes due to our assumptions.

This set of coupled equations can be written conveniently in matrix form
\begin{align}
  \begin{pmatrix} \partial_{tt} \phi \\ \partial_{tt} \psi \end{pmatrix} = 
  \begin{pmatrix} (c_2+c_3)/(2\rho_{\rm rot}) & -\chi_3/(4 \rho_{\rm rot}) \\ 
    -\chi_3/\rho & \mu/\rho \end{pmatrix}
  \begin{pmatrix} \Delta \phi \\ \Delta \psi \end{pmatrix}
  \label{eqn_tt2}
\end{align}  
where we used the symbol $\Delta$ for the 2d Laplacian. Let us denote the matrix on the right-hand side of~(\ref{eqn_tt2}) by ${\bf M}$. We note that the entries of ${\bf M}$ are all constants. Clearly, in the limit of no coupling $\chi_1 \rightarrow 0$, the waves do not interact and their respective waves speeds are given by $\sqrt{(c_2+c_3)/(2\rho_{\rm rot})}$ and $\sqrt{\mu/\rho}$, respectively, which is the expected result. 

Let us denote by ${\bf S}$ the matrix of eigenvectors of ${\bf M}$. Then we can diagonalise the coupled system~(\ref{eqn_tt2}) by writing
\begin{align}
  {\bf S}^{-1} \begin{pmatrix} \partial_{tt} \phi \\ \partial_{tt} \psi \end{pmatrix} = 
  ({\bf S}^{-1} {\bf M} {\bf S}) {\bf S}^{-1}\begin{pmatrix} \Delta \phi \\ \Delta \psi \end{pmatrix}
  \label{eqn_tt3}
\end{align}
where by construction $({\bf S}^{-1} {\bf M} {\bf S})$ is a diagonal matrix whose entries are the eigenvalues of ${\bf M}$. Thus, by linearly combining the elastic and rotational waves one finds two coupled waves which behave like ordinary plane waves. The explicit form of these solutions is quite involved, therefore we will only state explicitly the wave speeds of the coupled waves. They are given by
\begin{align}
  v^2_{\pm} = \frac{1}{2} \biggl(\frac{(c_2+c_3)}{\rho_{\rm rot}}+\frac{\mu}{\rho} \pm 
  \sqrt{\Bigl(\frac{(c_2+c_3)}{2\rho_{\rm rot}}-\frac{\mu}{\rho}\Bigr)^2+\frac{\chi_3^2}{\rho_{\rm rot}\,\rho}}\biggr).
  \label{eqn_tt4}
\end{align}
It is interesting to note that 
\begin{align}
  v^2_{+} + v^2_{-} = \frac{(c_2+c_3)}{2\rho_{\rm rot}} + \frac{\mu}{\rho}
  \label{eqn_tt5}
\end{align}
so that we can conclude that the sum of the wave speeds of the two coupled waves equals the sum of the wave speeds of the original uncoupled waves. This is expected as energy in the medium is conserved. Therefore, we can interpret this interaction by saying that one of the waves is speeding up while the other one slows down by an equal amount. 

The wave velocities~(\ref{eqn_tt4}) become most insightful if we consider $\chi_3$ to be much smaller than one and perform an expansion about $\chi_3 = 0$. In this case, we find
\begin{align}
  v^2_{+} &= \frac{(c_2+c_3)}{2\rho_{\rm rot}} + \frac{1}{4}\frac{\chi_1^2}{\rho_{\rm rot}\,\rho} \Bigl(\frac{(c_2+c_3)}{2\rho_{\rm rot}}-\frac{\mu}{\rho}\Bigr)^{-1} + O(\chi_1^2),
  \label{eqn_tt6}\\
  v^2_{-} &= \frac{\mu}{\rho} - \frac{1}{4}\frac{\chi_1^2}{\rho_{\rm rot}\,\rho} \Bigl(\frac{(c_2+c_3)}{2\rho_{\rm rot}}-\frac{\mu}{\rho}\Bigr)^{-1} + O(\chi_1^2).
  \label{eqn_tt7}
\end{align}
Let us denote $\tilde\chi = \chi_3^2/(4 \rho_{\rm rot}\,\rho)$ and introduce the notation $v_{\rm rot} = \sqrt{(c_2+c_3)/(2\rho_{\rm rot}})$, $v_{\rm elas} = \sqrt{\mu/\rho}$. This allows the following rewriting
\begin{align}
  v^2_{+} &= v_{\rm rot}^2 + \frac{\tilde\chi}{v_{\rm rot}^2 - v_{\rm elas}^2} + O(\tilde\chi^2),
  \label{eqn_tt8} \\
  v^2_{-} &= v_{\rm elas}^2 - \frac{\tilde\chi}{v_{\rm rot}^2 - v_{\rm elas}^2} + O(\tilde\chi^2).
  \label{eqn_tt9}
\end{align}
We can now provide a neat interpretation for the coupled waves. The $+$-wave corresponds to the rotational waves in the limit $\chi_3 \rightarrow 0$. If $v_{\rm rot} < v_{\rm elas}$ the coupling of the two waves has the effect of slowing down the rotational wave and speeding up the elastic waves which corresponds to the $-$-wave. In the other case when $v_{\rm rot} > v_{\rm elas}$, it is the rotational waves which becomes faster and the elastic waves which slows down. This means that whichever wave travels faster, before any couplings are introduced, will get even more energetic due to the couplings and vice versa. 

It should be noted that the case $v_{\rm rot} = v_{\rm elas}$ does not result in any problems. The divergent denominators in~(\ref{eqn_tt8}) and~(\ref{eqn_tt9}) are an artifact of the expansion about $\chi_1 = 0$. Equation~(\ref{eqn_tt4}) is well defined in this case and the respective wave speeds depend already linearly on $\chi_3$ as the first term in the square root vanishes.

\subsection{Longitudinal--Longitudinal coupling}

In analogy with the previous subsection we will now study the case where both waves are longitudinal. In this case we assume $\omega_x \equiv 0$, $\omega_y \equiv 0$, $\omega_z = \phi(z,t)$ and $u_x \equiv 0$, $u_y \equiv 0$, $u_z = \psi(z,t)$ so that $\mathrm{curl} \vect{\omega} = 0$ and $\mathrm{curl} \vect{u} = 0$. As before, four of the six equations are identically satisfied and we are left with the following two equations
\begin{align}
  \frac{1}{3}(c_1 + 2c_3) \partial_{zz} \phi + \frac{1}{6}(3\chi_1 -2\chi_3) \partial_{zz} \psi - \rho_{\rm rot} \partial_{tt} \phi &= 0
  \label{eqn_ll1}\\
  (\lambda + 2\mu) \partial_{zz} \psi + \frac{2}{3}(3\chi_1 -2\chi_3) \partial_{zz} \phi - \rho \partial_{tt} \psi &= 0
  \label{eqn_ll2}
\end{align}
which can be written as
\begin{align}
  \begin{pmatrix} \partial_{tt} \phi \\ \partial_{tt} \psi \end{pmatrix} = 
  \begin{pmatrix} (c_1+2c_3)/(3\rho_{\rm rot}) & (3\chi_1 -2\chi_3)/(6 \rho_{\rm rot}) \\ 
    (3\chi_1 -2\chi_3)/(3\rho) & (\lambda+2\mu)/\rho \end{pmatrix}
  \begin{pmatrix} \partial_{zz} \phi \\ \partial_{zz} \psi \end{pmatrix}
  \label{eqn_ll3}.
\end{align}  
As in the above, suitable linear combinations of the two elastic waves give plane wave. Their wave speeds are determined by the eigenvalues of the matrix~(\ref{eqn_ll3}) which are given by
\begin{multline}
  v^2_{\pm} = \frac{1}{6} \biggl(\frac{(c_1+2c_3)}{\rho_{\rm rot}} + \frac{3(\lambda+2\mu)}{\rho} \\\pm \sqrt{\Bigl(\frac{(c_1+2c_3)}{\rho_{\rm rot}}-\frac{3(\lambda+2\mu)}{\rho}\Bigr)^2 + \frac{4(3\chi_1 -2\chi_3)^2}{\rho_{\rm rot}\,\rho}} \biggr).
  \label{eqn_ll4}
\end{multline}
Inspection of the last term in the square-root of~(\ref{eqn_ll4}) shows that there exists a particular parameter choice $\chi_1 = 2\chi_2/3$ for which the waves speeds are unchanged. In this case, both coupling terms in the energy functional~(\ref{eqn_int}) take the same form under the assumption that both waves are longitudinal. Therefore, this cancellation is not unexpected.

\begin{figure}[!ht]
\begin{center}
\includegraphics[width=.15\textwidth]{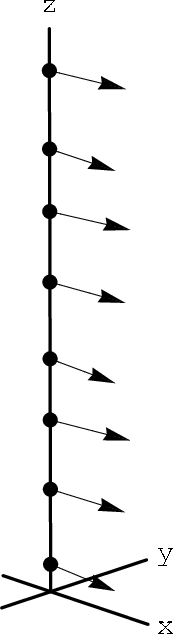}
\hfill
\includegraphics[width=.15\textwidth]{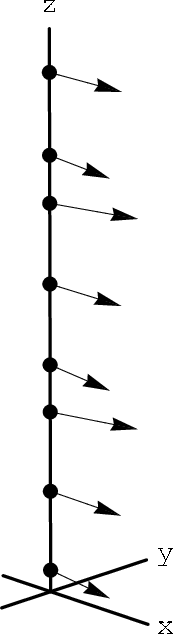}
\hfill
\includegraphics[width=.15\textwidth]{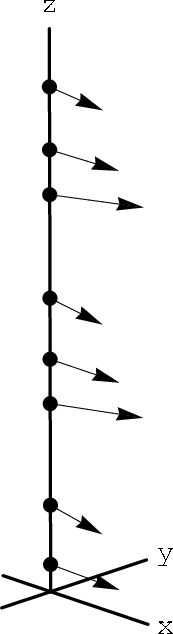}
\hfill
\includegraphics[width=.15\textwidth]{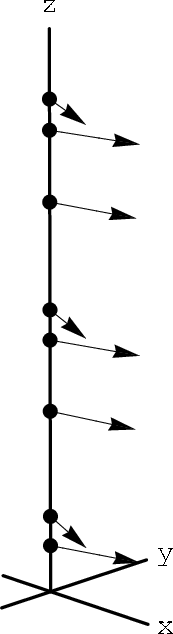}
\hfill
\includegraphics[width=.15\textwidth]{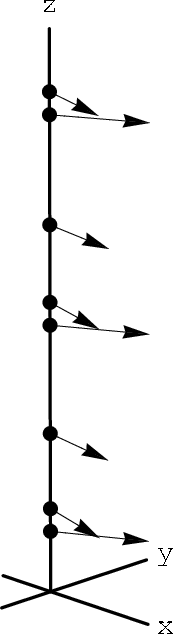}\\[2ex]
\includegraphics[width=.15\textwidth]{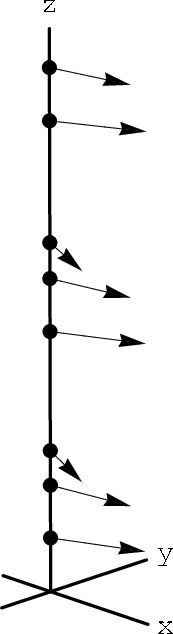}
\hfill
\includegraphics[width=.15\textwidth]{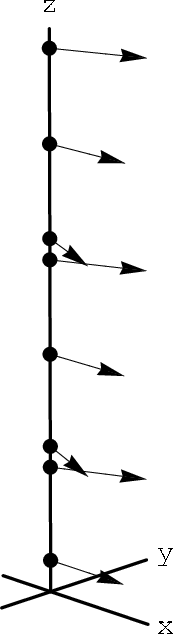}
\hfill
\includegraphics[width=.15\textwidth]{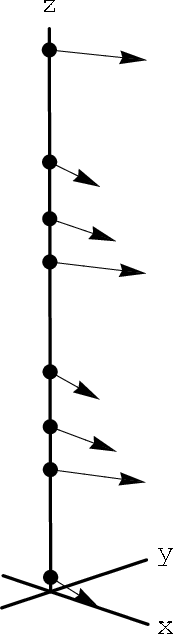}
\hfill
\includegraphics[width=.15\textwidth]{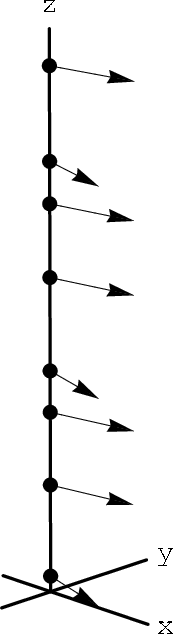}
\hfill
\includegraphics[width=.15\textwidth]{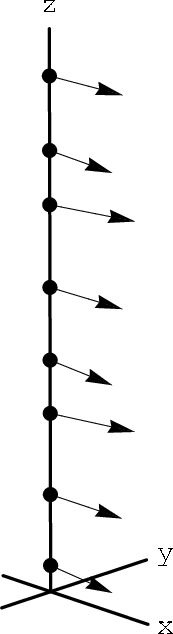}\\[2ex]
\caption{This figure visualises the propagation of the rotational and the elastic wave. Both waves are assumed to be longitudinal. Dots represent material points and the distance between point indicate the displacement. The arrows represent the rotation of the material points. Panel 1 shows the material points approximately equidistant and the arrows pointing roughly in the same direction. As time progresses, we see the displacement changing and also the arrows moving. Panels 5 and 6 represent the `maximum' of the wave, materials points are fully displaced and have rotated significantly. Panel 10 shows the end of the phase, points are roughly equidistant again and the arrows are pointing in the same direction.} 
\label{fig2}
\end{center}
\end{figure}

Figure~\ref{fig2} shows one possible visualisation of the rotational and the elastic wave for this case. We show one period of oscillation, one can see that different periodic motion exists for the rotational and also for the elastic wave.

Let us assume that the coupling $(3\chi_1 -2\chi_3)$ is much smaller than one and let us denote $v_{\rm rot} = \sqrt{(c_1+2c_3)/(3\rho_{\rm rot}})$ and $v_{\rm elas} = \sqrt{(\lambda+2\mu)/\rho}$. Then, we find
\begin{align}
  v^2_{+} &= v_{\rm rot}^2 + \frac{\tilde\chi}{v_{\rm rot}^2 - v_{\rm elas}^2} + O(\tilde\chi^2),
  \label{eqn_ll5} \\
  v^2_{-} &= v_{\rm elas}^2 - \frac{\tilde\chi}{v_{\rm rot}^2 - v_{\rm elas}^2} + O(\tilde\chi^2)
  \label{eqn_ll6}
\end{align}
where we also used $\tilde\chi = (3\chi_1 -2\chi_3)^2/(9 \rho_{\rm rot}\,\rho)$. The effects of the coupling in the case of longitudinal waves is very similar to that of transversal waves. As in the above we find that whichever wave travelled faster before any couplings were introduced will have its wave speed increased. 

\subsection{Transversal--Longitudinal coupling}

The last coupling we are considering is the case where the rotational wave is transversal and the elastic displacement wave is longitudinal. We assume $\omega_x \equiv 0$, $\omega_y \equiv 0$, $\omega_z = \phi(x,y,t)$ and $u_x \equiv 0$, $u_y \equiv 0$, $u_z = \psi(z,t)$ so that $\mathrm{div} \vect{\omega} = 0$ and $\mathrm{curl} \vect{u} = 0$. Unlike in the previous two cases, only two of the six equations are identically satisfied, and we are thus left with four equations. Of these four equations, two correspond to propagation equations (the coupled wave equations), while the remaining two equations impose strong constraints on the solutions. However, for the specific choice of coupling constants $\chi_1 = -\chi_3/3$, these equations are identically satisfied and place no additional restrictions on the possible solutions. In this case, the remaining two equations are given by 
\begin{align}
  (c_2+c_3) (\partial_{xx} + \partial_{yy}) \phi + \frac{3}{2} \chi_2 \partial_{zz} \psi - \rho_{\rm rot} \partial_{tt} \phi &= 0 \\
  (\mu+2\lambda) \partial_{zz} \psi + 3\chi_2 (\partial_{xx} + \partial_{yy}) \phi - \rho \partial_{tt} \psi &= 0,
  \label{eqn_tl1}
\end{align}
which, as before, can be written in the convenient form
\begin{align}
  \begin{pmatrix} \partial_{tt} \phi \\ \partial_{tt} \psi \end{pmatrix} = 
  \begin{pmatrix} (c_2+c_3)/(2\rho_{\rm rot}) & -\chi_3/(2 \rho_{\rm rot}) \\ 
    -\chi_3/\rho & (\lambda+2\mu)/\rho \end{pmatrix}
  \begin{pmatrix} (\partial_{xx} + \partial_{yy}) \phi \\ \partial_{zz} \psi \end{pmatrix}
  \label{eqn_tl3}.
\end{align}
As in the previous couplings, we can diagonalise the problem and identify two waves which in the limit of no coupling will reduce to the rotational and elastic waves, respectively. One easily verifies that it is again the faster wave which acquires additional energy due to the coupling. The most interesting aspect of this coupling is the necessity to impose a condition on the two coupling constants $\chi_1$ and $\chi_3$ in order to satisfy to additional equations. 

The Longitudinal--Transversal coupling case is similar. In this case the coupling depends on the constants $\chi_1$ and $\chi_3$ but the additional constraint equations force $\chi_3=0$. This leads to unchanged wave velocities, and we could not find simple interacting wave type solutions. 

\section{Conclusions and Discussions}

A new point of view was introduced to model rotational elasticity in a nonlinear setting using orthogonal matrices as the unknown variables. This follows the approach successfully taken in~\cite{Boehmer:2012} where soliton like solutions were found. We showed how this theory can be related to classical elasticity and identified parameter ranges where the rotational medium would correspond to an auxetic material. After identifying the most general energy functional of this model, two types of plane wave solutions were constructed analytically, they are solutions of the nonlinear Euler-Lagrange equations. These waves correspond to transversal rotational waves and to longitudinal rotational waves. Similar waves were found in~\cite{Boehmer:2010} using similar ideas but a different model. It is interesting that in our rotational setting the transversal wave velocity can be greater than the longitudinal wave velocity. 

We coupled the rotational elasticity to linear elasticity describing displacements and searched for plane wave solutions of the coupled model in the fully non-linear setting. We were able to find various types of solutions by assuming that the rotational and displacement waves were either transversal or longitudinal, respectively. Figure~\ref{fig2} visualises the plane wave solutions in the case where both waves are longitudinal. It will be interesting to study the proposed model further and in particular to investigate the existence of soliton like solutions in the presence of couplings between rotations and displacements.

We would like to briefly refer to~\cite{Neff:2009} where the authors call for deeper differential geometric insight to motivate a conformally invariant curvature term. Our approach is entirely based on differential geometry, the energy functional~(\ref{neweqn}) is the most general which can be constructed from the contortion tensor. It happens that one of these three irreducible parts is invariant under conformal transformations, it is precisely that term which is studied in the context the Dirac equation in~\cite{Burnett:2008bx,Chervova:2010,Burnett:2012} and also in the context of Cosserat elasticity in~\cite{Neff:2009,Neff:2010,Neff:2010b}. It is most surprising that the same object features in two such distinct fields of research. It appears to be worthy to study conformally invariant geometries in three dimension in more detail. Using an arbitrarily deformed medium and studying the behaviour of the Cotton tensor (the vanishing of the Cotton tensor is the necessary and sufficient condition for the manifold to be conformally flat, similarly to the Weyl tensor in dimension $\geq 4$) might shed some light into these interesting issues. In~\cite{Boehmer:2012} where soliton like solutions were found, it was shown that these solutions have a topological soliton. There seems to be a hidden connection between Cosserat elasticity and differential geometry.

\section*{Acknowledgement(s)}

We would like to thank Matias Dahl, Robert Downes, Friedrich Hehl, Patrizio Neff, Yuri Obukhov and Dmitri Vassiliev for the fruitful discussions and comments on the manuscript.

This research received no specific grant from any funding agency in the public, commercial, or not-for-profit sectors.

\providecommand{\bysame}{\leavevmode\hbox to3em{\hrulefill}\thinspace}


\begin{thebibliography}{99}

\bibitem{Ball:2008}
  J.~M. Ball and A. Zarnescu, 
  \emph{{Orientable and Non-Orientable Line Field Models for Uniaxial Nematic Liquid Crystals}}, 
  Mol. Cryst. Liq. Cryst. \textbf{495} (2008), 221--233.

\bibitem{Boehmer:2010}
  C.~G. B{\"o}hmer, R.~Downes, and D.~Vassiliev, 
  \emph{{Rotational Elasticity}}, 
  Q. J. Mech. Appl. Math. \textbf{64} (2011), 415--439.

\bibitem{Boehmer:2012}
  C.~G. B{\"o}hmer and Yu.~N.~Obukhov, 
  \emph{{A gauge theoretic approach to elasticity with microrotations}}, Proc. R. Soc. A.
  \textbf{468} (2012), 1391--1407.

\bibitem{Burnett:2008bx}
  J. Burnett, O. Chervova, and D. Vassiliev, 
  \emph{{Dirac equation as a special case of Cosserat elasticity}}, 
  {Analysis, Partial Differential Equations and Applications -- The Vladimir Maz'ya Anniversary Volume} 
  (Basel) (A. Cialdea, F. Lanzara, and P.~E. Ricci, eds.), 
  Operator Theory: Advances and Applications, vol. 193, 
  Birkhauser Verlag, 2008, pp.~15--29.

\bibitem{Burnett:2012}
  J. Burnett and D. Vassiliev,
  \emph{{Modeling the electron with Cosserat elasticity}},
  Mathematika \textbf{58} (2012), 349--370.

\bibitem{capriz}
  G. Capriz, \emph{{Continua with microstructure}}, 
  Springer Tracts in Natural Philosophy \textbf{35}, 
  Springer, Berlin, 1989.

\bibitem{Chervova:2010}
  O. Chervova and D. Vassiliev,
  \emph{{The stationary Weyl equation and Cosserat elasticity}},
  J. Phys. A: Math. Theor. \textbf{43} (2010) 335203.

\bibitem{Cosserat:1909}
  E. Cosserat and F. Cosserat, 
  \emph{{Th\'eorie des corps d'eformables}},
  Librairie Scientifique A.~Hermann et fils, Paris, 1909, 
  (Reprinted by Cornell University Library).

\bibitem{Del:2013}
  D.~H. Delphenich, 
  \emph{{The use of the teleparallelism connection in continuum mechanics}}, 
  arXiv:1305.3477 [gr-qc], 2013.

\bibitem{dysz}
  J. Dyszlewicz, 
  \emph{{Micropolar theory of elasticity}}, 
  Lecture Notes in Applied and Computational Mechanics, v. 15, 
  Springer, Berlin, 2004.

\bibitem{Ericksen:1962b}
  J.~L. Ericksen, 
  \emph{{Hydrostatic theory of liquid crystals}}, 
  Arch. Rational Mech. Anal. \textbf{9} (1962), 379--394.

\bibitem{Ericksen:1962a}
  \bysame, 
  \emph{{Kinematics of macromolecules}}, 
  Arch. Rational Mech. Anal. \textbf{9} (1962), 1--8.

\bibitem{Ericksen:1967}
  \bysame, 
  \emph{{Twisting of liquid crystals}}, 
  J. Fluid Mech. \textbf{27} (1967), 59--64.

\bibitem{Ericksen:1968}
  \bysame, 
  \emph{{Twist waves in liquid crystals}}, 
  Q. J. Mech. Appl. Math. \textbf{21} (1968), 463--465.

\bibitem{Eringen:1964a}
  A.~C. Eringen and E.~S. Suhubi, 
  \emph{{Nonlinear theory of simple microelastic solids I}}, 
  Int. J. Eng. Sci. \textbf{2} (1964), 189--204.

\bibitem{Eringen:1964b}
  \bysame, 
  \emph{{Nonlinear theory of simple microelastic solids II}}, 
  Int. J. Eng. Sci. \textbf{2} (1964), 389--404.

\bibitem{Eringen99}
  \bysame, 
  \emph{{Microcontinuum field theories: I. Foundations and solids}},
  Springer, New York, 1999.

\bibitem{Evans:1998}
  L.~C.~Evans, 
  \emph{Partial differential equations}, 
  American Mathematical Society, Providence, 1998.

\bibitem{Green:1964}
  A.~E. Green, 
  \emph{{Multipolar continuum mechanics}}, 
  Arch. Rational Mech. Anal. \textbf{17} (1964), 113--147.

\bibitem{gg1}
  G. Grioli, 
  \emph{{Elasticit\`a asimmetrica}}, 
  Ann. Mat. pura e Appl. \textbf{50} (1960), 389--417. 

\bibitem{gg2}
  \bysame, \emph{{Contributo per una formulazione di tipo integrale della 
  meccanica dei continui di Cosserat}}, Ann. Mat. pura e Appl. 
  \textbf{111} (1976), 175--183.

\bibitem{gg3}
  \bysame, \emph{{Microstructures as a refinement of Cauchy theory. Problems 
of physical concreteness}}, Continuum Mechanics and Thermodynamics
  \textbf{15} (2003), 441--450. 

\bibitem{mag}
  F.W. Hehl, J.D. McCrea, E.W. Mielke, and Y. Ne'eman, 
  \emph{{Metric-affine gauge theory of gravity: Field equations, Noether identities, world spinors, and breaking of dilation invariance}}, 
  Phys. Repts. \textbf{258} (1995), 1--171.

\bibitem{fhyo}
  F.~W.~Hehl and Yu.~N.~Obukhov, 
  \text{{\'Elie Cartan's torsion in geometry and in field theory, an essay}}, 
  Ann. Fond. Louis Broglie \textbf{32} (2007), 157--194.

\bibitem{Neff:2010}
  J.~Jeong and P.~Neff, 
  \emph{{Existence, Uniqueness and Stability in Linear Cosserat Elasticity for Weakest Curvature Conditions}}, 
  Math. Mech. Solids \textbf{15} (2010), 78--95.

\bibitem{Jeong:2009}
  J.~Jeong, H.~Ram\'{e}zani, I.~M\"unch, and P.~Neff, 
  \emph{{A numerical study for linear isotropic Cosserat elasticity with conformally invariant curvature}}, 
  Z. Angew. Math. Mech. \textbf{89} (2009), 552--269.

\bibitem{Kat1}
  M.~O.~Katanaev and I.~V.~Volovich, 
  \emph{{Theory of defects in solids and three-dimensional gravity}}, 
  Ann. Phys. (USA) \textbf{216} (1992), 1--28. 

\bibitem{Kat2}
  M.~O.~Katanaev, 
  \emph{{Wedge dislocation in the geometric theory of defects}},
  Theor. Math. Phys. \textbf{135} (2003), 733--744.

\bibitem{Lakes:1987}
  R.~S. Lakes, \emph{{Foam structures with a negative Poisson's ratio}}, Science
  \textbf{235} (1987), 1038--1040.

\bibitem{Lazar:2010}
  M.~Lazar, 
  \emph{{On the fundamentals of the three-dimensional translation gauge theory of dislocations}}, 
  Mathematics and Mechanics of Solids
  \textbf{16} (2011), 253--264.

\bibitem{Lazar:2009ga}
  M.~Lazar and F.~W. Hehl, 
  \emph{{Cartan's spiral staircase in physics and, in particular, in the gauge theory of dislocations}}, 
  Found. Phys. \textbf{40} (2010), 1298--1325.

\bibitem{Love:1944}
  A.~E.~H. Love, 
  \emph{A treatise on the mathematical theory of elasticity},
  Dover, New York, 1944.

\bibitem{Madeo:2013a}
  A.~Madeo, P.~Neff, I.-D.~Ghiba, L.~Placidi, G.~Rosi,
  \emph{{Band gaps in the relaxed linear micromorphic continuum}},
  preprint.

\bibitem{Madeo:2013b}
  \bysame,
  \emph{{Wave propagation in relaxed micromorphic continua: modelling metamaterials with frequency band-gaps}},
  Contin. Mech. Thermodyn. (submitted).

\bibitem{Majumdar:2010}
  A.~Majumdar and A.~Zarnescu, 
  \emph{{Landau-de Gennes Theory of Nematic Liquid Crystals: the Oseen-Frank Limit and Beyond}}, 
  Arch. Rational Mech. Anal. \textbf{196} (2010), 227--280.

\bibitem{Mindlin:1964}
  R.~D.~Mindlin, 
  \emph{{Microstructure in linear elasticity}}, 
  Arch. Rational Mech. Anal. \textbf{16} (1964), 51--78.

\bibitem{Muskhelishvili:1953}
  N.~I. Muskhelishvili, 
  \emph{{Some basic problems of the mathematical theory of elasticity}}, 
  P. Noordhoff, Groningen, Holland, 1953.

\bibitem{na1}
  D.~Natroshvili, L.~Giorgashvili and I.~G.~Stratis, 
  \emph{{Mathematical problems of the theory of elasticity of chiral materials}}, 
  Applied Mathematics, Informatics and Mechanics \textbf{8} (2003), 47--103.

\bibitem{na2}
  \bysame, 
  \emph{{Representation formulae of general solutions in the theory of hemitropic elasticity}}, 
   Q. J. Mech. Appl. Math. \textbf{59} (2006), 451--474.

\bibitem{na3}
  D.~Natroshvili, R.~Gachechiladze, A.~Gachechiladze and I.~G.~Stratis, 
  \emph{{Transmission problems in the theory of elastic hemitropic materials}}, 
  Applicable Analysis \textbf{86} (2007), 1463--1508.

\bibitem{Neff:2006a}
  P.~Neff, 
  \emph{{Existence of minimizers for a finite-strain micromorphic elastic solid}}, 
  Proc. Roy. Soc. Edinb. A \textbf{136} (2006), 997--1012.

\bibitem{Neff:2006b}
  P.~Neff,
  \emph{{A finite-strain elastic-plastic Cosserat theory for polycrystals with grain rotations}},
  Internat. J. Engrg. Sci. \textbf{44} (2006), 574--594.

\bibitem{Neff:2006c}
  P.~Neff, 
  \emph{{The Cosserat couple modulus for continuous solids is zero viz the linearized Cauchy-stress tensor is symmetric}},
  Z. Angew. Math. Mech. \textbf{86} (2006), 892--912.

\bibitem{Neff:2007a}
  P.~Neff,
  \emph{{A geometrically exact planar Cosserat shell-model with microstructure: existence of minimizers for zero Cosserat couple modulus}},
  Math. Models Methods Appl. Sci. \textbf{17} (2007), 363--392.

\bibitem{Neff:2008}
  P.~Neff,
  \emph{{Remarks on invariant modelling in finite strain gradient plasticity}},
  Technische Mechanik \textbf{28} (2008), 13--21.

\bibitem{Neff:2009ca}
  P.~Neff, K.~Che\l{}mi{/'n}ski, H.-D.~Alber, 
  \emph{{Notes on strain gradient plasticity: finite strain covariant modelling and global existence in the infinitesimal rate-independent case}},
  Math. Models Methods Appl. Sci. \textbf{19} (2009), 307--346.

\bibitem{Neff:2007b}
  P.~Neff and S.~Forest,
  \emph{{A Geometrically Exact Micromorphic Model for Elastic Metallic Foams Accounting for Affine Microstructure. Modelling, Existence of Minimizers, Identification of Moduli and Computational Results}},
  J. Elasticity \textbf{87} (2007), 239--276.

\bibitem{Neff:2009}
  P.~Neff and J.~Jeong, 
  \emph{{A new paradigm: the linear isotropic Cosserat model with conformally invariant curvature energy}}, 
  Z. Angew. Math. Mech. \textbf{89} (2009), 107--122.

\bibitem{Neff:2010a}
  P.~Neff, J.~Jeong, and A.~Fischle,
  \emph{{Stable identification of linear isotropic Cosserat parameters: bounded stiffness in bending and torsion implies conformal invariance of curvature}},
  Acta Mechanica \textbf{211} (2010), 237--249.

\bibitem{Neff:2010b}
  P.~Neff, J.~Jeong, I.~M\"unch, and H.~Ram\'{e}zani,
  \emph{{Linear Cosserat Elasticity, Conformal Curvature and Bounded Stiffness}}, 
  In: Mechanics of Generalized Continua, G.~A.~Maugin, A.~V.~Metrikine (eds.),
  Advances in Mechanics and Mathematics \text{21} (2010), 55--63.

\bibitem{Neff:2008n}
  P.~Neff and I.~M\"unch,
  \emph{{Curl bounds Grad on SO(3)}},
  ESAIM: Control, Optimisation and Calculus of Variations \textbf{14} (2008), 148--159.

\bibitem{Neff:2009b}
  P.~Neff and I.~M\"unch,
  \emph{{Simple shear in nonlinear Cosserat elasticity: bifurcation and induced microstructure}},
  Contin. Mech. Thermodyn. \textbf{21} (2009), 195--221.

\bibitem{Neff:2013a}
  P.~Neff, I.-D.~Ghiba, A.~Madeo, L.~Placidi, G.~Rosi,
  \emph{{A unifying perspective: the relaxed linear micromorphic continuum}},
  Contin. Mech. Thermodyn. (to appear).

\bibitem{nowacki}
  W. Nowacki, \emph{{Theory of asymmetric elasticity}}, 2nd ed., 
  Pergamon, Oxford, 1985.

\bibitem{Schaefer:1967}
  H.~Schaefer, \emph{{Das Cosserat-Kontinuum}}, 
  Z. Angew. Math. Mech. \textbf{47} (1967), 485--498.

\bibitem{sch2}
  \bysame, \emph{{Analysis der Motorfelder im Cosserat-Kontinuum}}, 
  Z. Angew. Math. Mech .\textbf{47} (1967), 319--328. 

\bibitem{sch3}
  \bysame, \emph{{Die Motorfelder des dreidimensionalen Cosserat-Kontinuums im Kalk\"ul der Differentialformen}}, 
  Lecture course No. 19, International Centre for Mechanical Sciences 
  (ICMS, Udine, 1970) 1--60. 

\bibitem{Skyrme:1961}
  T.~H.~R. Skyrme, 
  \emph{A non-linear field theory}, 
  Proc. Roy. Soc. A \textbf{260} (1961), 127--138.

\bibitem{Sokolnikoff:1956}
  I.~S.~Sokolnikoff, 
  \emph{{Mathematical Theory of Elasticity}}, 
  McGraw-Hill, New York, 1956.

\bibitem{Toupin:1962}
  R.~A. Toupin, 
  \emph{{Elastic materials with couple-stresses}}, 
  Arch. Rat. Mech. Anal. \textbf{11} (1962), 385--414.

\bibitem{Toupin:1964}
  \bysame, 
  \emph{{Theories of elasticity with couple-stress}}, 
  Arch. Rational Mech. Anal. \textbf{17} (1964), 85--112.

\bibitem{TT}
  C. Truesdell and R. Toupin, 
  \emph{{The classical field theories. Encyclopedia of Physics}}, 
  Vol. III/1., Berlin-G\"ottingen-Heidelberg, Springer, 1960.
   
\bibitem{Whittaker_VolI}
  E.~Whittaker, 
  \emph{{A History of the Theories of Aether and Electricity}},
  Thomas Nelson and Sons, London, United Kingdom, 1951.

\bibitem{Yavari:2012}
  A.~Yavari and A.~Goriely,
  \emph{{Riemann-Cartan geometry of nonlinear dislocation mechanics}},
  Arch. Rational Mech. Anal. \textbf{205} (2012), 59--118.

\bibitem{Yavari:2012a}
  \bysame, 
  \emph{{Riemann-Cartan geometry of nonlinear disclination mechanics}}, 
  Mathematics and Mechanics of Solids 
  \textbf{18} (2013) 91--102.

\end{thebibliography}
\end{document}